\documentclass[preprint]{ptephy_v1}

\preprintnumber{1205.1835 hep-ph}
\usepackage{hyperref}

\newcommand{\req}[1]{Eq.\,(\ref{#1})} 

\begin{document}

\title{A Cusp in QED at $g=2$}


\author{Johann Rafelski}
\affil{Department of Physics, The University of Arizona, Tucson, AZ 85721, USA 
\\ \vskip 0.5cm} 
 
\author{Lance Labun}
\affil{Department of Physics, The University of Texas, Austin, TX 78712, USA}

\begin{abstract}%
We explore nonperturbative properties of QED allowing a gyromagnetic ratio $g\ne g_{\rm D}\equiv 2$. We study the effective action $V_{\mathrm{eff}}$ for an arbitrarily strong constant and homogeneous field. Using the external field method, we find a cusp as a function of the gyromagnetic factor $g$ in: a) The QED $b_0$-renormalization group coefficient; b) A subclass of light-light scattering coefficients obtained in the long wavelength limit expansion. We recognize possibility of asymptotic freedom in an Abelian theory for certain domains of $g$.
\end{abstract}

\subjectindex{PTEP: B30}

\maketitle

\noindent {\bf 1. Motivation:}\\
\noindent No known particle has exactly the Dirac value $g_\mathrm{ D}\equiv 2$ of the gyromagnetic ratio $g$. Determination of the higher order vacuum fluctuation correction to $g_\mathrm{ D}$ provides the most precise test of perturbative QED (D-QED)~\cite{Kinoshita:2005sm}. However, even the point-like electron (or muon or tauon) to start with have $g\ne 2$ due to modifications introduced by electromagnetic interactions with particles not recognized as present in the QED framework - these, for example, are EM interactions with quark fluctuations in the vacuum. However, should $g=g_\mathrm{D}$ be a singular point, a D-QED perturbative expansion is not appropriate. The aims of this work are to recognize a singularity at $g=g_\mathrm{ D}$, to study the nature of this singularity, and to lay foundation for a theoretical framework allowing exploration of $|g|>g_\mathrm{ D}$ domain.

To achieve our goals we consider the extension to $g\ne 2$ based on the renormalizable dimension-4 action~\cite{VaqueraAraujo:2012qa,AngelesMartinez:2011nt}.  We study the vacuum properties in the presence of external constant and homogeneous electromagnetic fields, integrating out fluctuations of spin-1/2 particles with $g\ne g_\mathrm{ D}$. The resulting effective potential $V_{\mathrm{eff}}$ is a generalization of the Heisenberg-Euler-Schwinger (HES) effective action~\cite{Heisenberg:1935qt,Weisskopf,Schwinger:1951nm,Reuter:1996zm,Dunne:2004nc} to arbitrary value of $g$. The result is regular for all $|g|\le g_\mathrm{ D}$~\cite{Labun:2012jf}. 

For $|g|> g_\mathrm{ D}$ the HES proper time effective action becomes non-integrable. We propose a natural analytic extension of the  proper time integrand to all values $|g|> g_\mathrm{ D}$, which shows that $g=g_\mathrm{ D}$ (and other periodic recurrent values) is a cusp point as a function of~$g$. This extension resolves the known difficulties in the theoretical framework of $g\ne g_\mathrm{ D}$ theories~\cite{Veltman:1997am}. Moreover, considering the beta-function coefficients, we demonstrate the domains for which asymptotic freedom arises as a function of $g$. 

Since we do not continue analytically the actual effective potential $V_{\mathrm{eff}}$  but the  proper time integrand that yields this result, the mathematical exactness of the here presented work can be questioned. Therefore a confirmation of the analytic extension as proposed here was developed following ideas seen in the work of Heisenberg-Euler~\cite{Heisenberg:1935qt}, applying recently developed methods demonstrated in Ref.~\cite{Evans:2022fsu}. This leads to a confirmation of results obtained here in an ab-initio approach, which, along with further results, will be presented in a longer forthcoming publication~\cite{Evans:2022EB}.

\noindent{\bf 2. Introducing magnetic moment $\mathbf{|g|\ne g_\mathrm{ D}}$:}\\
\noindent One way to account for $ |g|\ne g_\mathrm{ D} $ is to complement the Dirac action with an incremental Pauli interaction term $\delta\!\mu\,(\vec \sigma\cdot \vec B+i\vec \alpha \cdot \vec E)=\delta\!\mu\,\sigma_{\alpha\beta}F^{\alpha\beta}/2$, where $\vec E, \vec B$ are the electromagnetic fields, $F^{\alpha\beta}$ the electromagnetic field strength tensor, $\sigma_{\alpha\beta}=(i/2)[\gamma_{\alpha},\gamma_{\beta}]$ with $\gamma_\alpha$ the usual Dirac matrices, and $\vec \sigma$, and $\vec \alpha=\gamma_5\vec\sigma$ are the Pauli-Dirac matrices. However, such incremental Pauli interaction is a dimension 5 operator, $[\,\overline\psi\sigma_{\alpha\beta}F^{\alpha\beta}\psi]=L^{-5}$. The coefficient $\delta\!\mu$ consequently has dimension length, which in the case of a composite particle such as the proton, is naturally related to the particle size. Therefore this Dirac-Pauli (DP) modification of the Dirac equation has been a popular and effective tool to describe to lowest order the magnetic moment dynamics of a composite particle of finite size, for example a proton.

For a point elementary particle such as the electron it is more appropriate to start for $|g|\neq g_\mathrm{D}$ by adding the full Pauli spin-interaction term to the Klein-Gordon action
\begin{equation}\label{EoMg}
{\cal L}=\bar\psi\left[\Pi^2-m^2-\frac{g}{2} \frac{e \sigma_{\alpha\beta}F^{\alpha\beta}}{2}\right]\psi,
\end{equation}
where $\Pi_\alpha=i\partial_\alpha+eA_\alpha$. Note that the dimension of the $\psi$ field is $[\psi]=L^{-1}$ and consequently the Pauli interaction is dimension 4. We refer to the study of QED based on~\req{EoMg} as $g$-QED, and the dynamical equation following from~\req{EoMg} as the Klein-Gordon-Pauli (KGP) equation. $g$-QED is the $s=1/2$ case in the study of particles of all spins in the Poincar\'e group framework initiated by Rarita and Schwinger~\cite{Rarita:1941mf, Napsuciale:2006wr}. For related developments see references in the introduction to Ref.\cite{VaqueraAraujo:2012qa}, and for detailed comparison between the KGP and DP methods see~\cite{Steinmetz:2018ryf}.

Since there are at least two distinct paths to introduce $g\ne 2$ corrections into relativistic particle dynamics, the question is in what sense these could be equivalent and if not, which of the two forms is appropriate for study of particle dynamics and/or vacuum structure and under what conditions:\\[0.1cm]
1) The DP approach, involving a di\-men\-sion-5 operator, requires new counter terms in each order. This, in our opinion, limits the DP approach to situations in which the physical particle properties are known and vacuum fluctuations need not be considered. Even so, we see in literature DP method applied to both vacuum fluctuation and the effective action evaluation in QED. see for example Refs.~\cite{PauliTerm, Diet78,Lav85, Ferrer:2015wca, Ferrer:2019xlr, Adorno:2021xvj}. \\[0.1cm]
2) In $g$-QED the magnetic moment remains point-like, $g\neq g_{\rm D}$ does not require a higher dimensioned operator. Therefore the quantum field theory requires a finite number of counter terms and is renormalizable~\cite{VaqueraAraujo:2012qa,AngelesMartinez:2011nt}; vacuum fluctuations can be considered in any perturbative order. \\[0.1cm]
3) It should be remembered that in $g$-QED an expansion around $g=2$ requires additional consideration since the natural expansion occurs around $g=0$. Properties of the KGP-originating non-perturbative effective action were considered for general spin in Ref.~\cite{Kruglov:2001dp}, but this work did not recognize the restricted validity domain of the perturbative approach, for spin-1/2 $-2\le g\le 2$ which arise due to convergence properties of the proper time Schwinger integral. \\[0.1cm]
4) Another study of quantum field amplitudes with an anomalous moment~\cite{Larkoski:2010am} also arrives at a second-order effective theory, but for a reduced two-component spinor. Given the derivation and properties of their effective theory, we believe that an exact relation between KGP and DP approaches can at best arise in an infinite order resummation in some specific applications.

Veltman has considered reduction of the number of dynamical components working in a two-component formulation. However, there are unresolved challenges~\cite{Veltman:1997am} in particular related to self-adjointness of the resulting spectrum and thus conservation of probability in temporal evolution. By individually characterizing states, we will present another resolution of this problem that works in presence of externally applied fields.

Considering that~\req{EoMg} is 2nd order in time and has four components, the number of dynamical degrees of freedom present in~\req{EoMg} is 8. That is, there are twice as many degrees of freedom as in usual Dirac theory. For the case $g=2$~\req{EoMg}, can be presented as the square of the operator $\gamma_5 D,\ D=\gamma_\alpha(i\partial_\alpha+eA^\alpha)-m$ and $\gamma_5=i\gamma^0\gamma^1\gamma^2\gamma^3, \gamma_5^2=1$ is the 5th Dirac matrix. This means that for $g=2$~\req{EoMg} comprises exact duplication of the Dirac degrees of freedom, the second set with opposite sign of mass $m$ which means with opposite association of the sign between magnetic moment and electric charge, see paragraphs below; for $g\ne 2$ one must search for a projection restricting the full Hilbert space to the physical states.

\noindent{\bf 3. Eigenvalue-sum periodicity as a function of $g$:}\\
\noindent We seek to identify the physics content of the 8 degrees of freedom and to separate the Hilbert space into two equal size parts that each individually comprises a complete set of states at a fixed given value of $g$. To do so, we consider the Landau-orbit spectrum of the operator in brackets in~\req{EoMg} in the presence of a constant magnetic field $\vec B$
\begin{equation}\label{spectrum}
E_n =\pm\sqrt{m^2+p_z^2+Q|e\vec B|\,[(2n\!+\!1) \mp g/2]},\quad Q=\pm 1,
\end{equation}
where $p_z$ is the one dimensional continuous momentum eigenvalue and $n$ is the Landau orbit quantum number. 

We have made explicit the presence of 8 eigenvalues for each value of $\vec B$, corresponding to all different possible choices of the three $\pm$-sign sets. There are the usual two roots in~\req{spectrum}, a known feature of relativistic dynamics also seen in the Landau spectrum of the Dirac equation where the negative energy states become positive energy antiparticle `hole' states. There is a new spectrum duplication related to two possible values of $Q$. This factor arises from two possible particle spin projections onto magnetic field, corresponding to the spin degeneracy.

\begin{figure}[t]
\begin{center}
\includegraphics[width=0.470\textwidth]{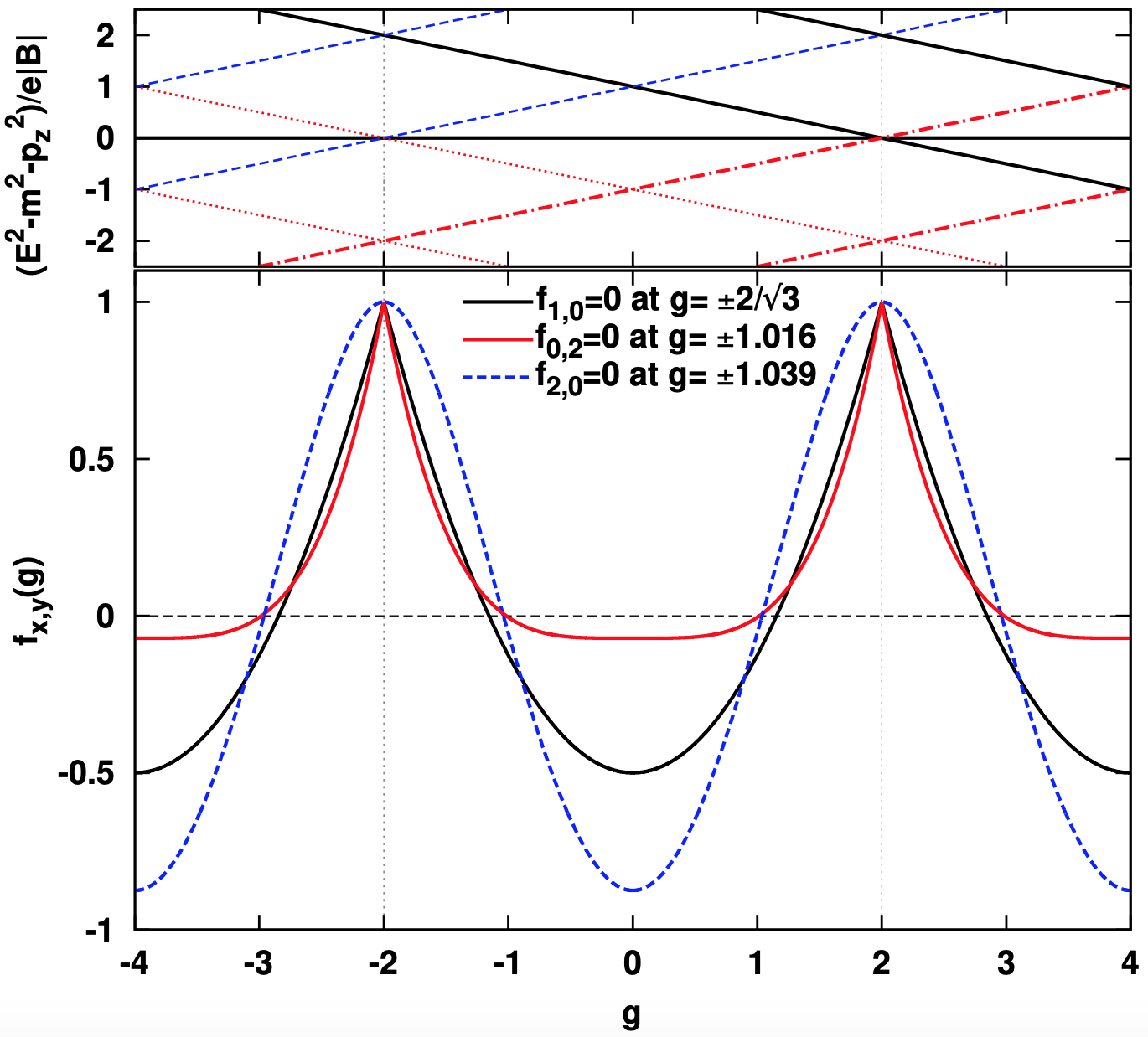}
\caption{Top: {Squared} eigenvalues~\req{spectrum} of KGP in magnetic field; {the solid and (blue) dashed lines are for $Q=+1$, and respectively $(-)$ and $(+)$ spin eigenvalue; the dotted and dash-dotted (red) lines are for $Q=-1$, and respectively $(+)$ and $(-)$ spin eigenvalue.} Bottom: coefficient functions: $f_{1,0}(g)$ as defined in~\req{betafunc} and $f_{0,2}$ and $f_{2,0}$ as defined in~\req{Veffexp}. Two full periods are shown. The values of $g$ where 
functions $f_{i,j}$ change sign
is indicated. \label{fig:Veffexpg}}
\end{center}
\end{figure}

To see how $Q$ can be restricted let us consider~\req{spectrum} in the form
\begin{equation}\label{spectrum1}
K=\frac{E_n^2 -m^2-p_z^2}{ |e\vec B|}=Q\,[(2n\!+\!1) \mp g/2],\quad Q=\pm 1.
\end{equation}
The quantity $K$ is shown in the top portion of figure \ref{fig:Veffexpg} as a function of $g$. We see that between $-2\le g\le 2$ there is an exact duplication of the spectrum corresponding to $Q=1$ and $Q= -1$. These are two sectors of the Hilbert space with the same physical content. The `squared' Dirac operator produces two eigenstate-space copies which can be separated in particular applications. Without restriction of generality the $Q=-1$ eigenvalues can be therefore omitted. Thus for $-2\le g\le 2$ the effective action is obtained by the usual procedure, and the results have already been presented~\cite{Labun:2012jf}.

For $|g|>2$, new and non-perturbative physics content arises for external fields of any strength, including arbitrarily weak. First we note that taking~\req{spectrum} expression at face value, naively some eigenstates could have $E^2<m^2$, which implies existence of bound localized states in the presence of a constant magnetic field. Such solutions are not required for completeness and would violate Lorentz symmetry; for these reasons, such states cannot be admitted in the spectrum. This situation differs from the $m^2+p_z^2\to 0$ limit, in which states having $K<0$ signal instabilities of the conventional vacuum state~\cite{Savvidy:1977as,Nielsen:1978rm}.

To compute the effective action we must define which states contribute to the physical spectral sum. The first step is to accomplish (like for the case $|g|\le 2$) separation of the Hilbert space into two sectors. We divide the states according to whether $K\geq 0$ or $K\leq 0$ and denote the respective sectors ${\cal K}^{\pm}$. The limit $K=0$ where two states coincide occurs at $g=2$ since the KGP operator can be written as the exact square of the Dirac operator. This situation recurs with the shift of $g$ by $4k, k\in \mathbb{Z}$. There is no change in the number of states in each of the Hilbert space sectors ${\cal K}^{\pm}$ as an equal number of single particle states is exchanged between both sectors.

The principle we use to determine which states enter the spectral sum is that there should be no localized bound states in a constant magnetic field. In the notation just introduced, we require $K\geq 0$ and the ${\cal K}^+$ sector be chosen as representing the physical spectrum. This is an extension from the regular case $|g|\leq 2$, where the usual procedure sums over the $Q=+1$ states and is equivalent to summing over the ${\cal K}^+$ state space. As $K\geq 0$ implies $E^2\geq m^2$, the physics is a continuous extension of the case $g=2$, for which it is proved that $E^2\geq m^2$ for arbitrary magnetic fields, i.e. there are no bound states~\cite{Gornicki:1987hv}. 

Looking far outside the principal domain $-2\le g\le 2$, we see that relativistic Landau eigenstates cross between ${\cal K}^\pm$ at each $g_k=2+ 4k, k\in \mathbb{Z}$. As the graphic representation top frame of Fig.\,\ref{fig:Veffexpg} shows, for each of the Hilbert space sectors ${\cal K}^{\pm}$ we have periodicity of the Landau levels a function of $g$. Therefore, the sum $\sum_{n}E_n$ over ${\cal K}^+$ leading to the real part of $V_{\mathrm{eff}}(\vec B^{\,2})$ is a periodic function of $g$, a result we will find explicitly. This periodicity does not apply to individual Landau eigenvalues as is seen in~\req{spectrum}. In computation of vacuum fluctuations the truncation of the Landau eigenstate $n$-sum to any finite value breaks the periodicity as well.

The choice of ${\cal K}^+$ as the physical state space has clear advantages and resolves the challenges encountered by Veltman~\cite{Veltman:1997am}: In addition to maintaining self-adjointness of the KGP system, it makes the quantum field theories based on semi-spaces ${\cal K}^{\pm}$ each individually unitary, because the number of states is conserved in transiting through the singular points e.g at $|g|=2$, and for $|g|>2$ we omit the localized solutions. Moreover, our proposal makes the spectrum and by extension the quantum theory a continuous and analytic extension from the domain $|g|\leq 2$. Our approach preserves translation invariance of the vacuum, which would be broken by any localized bound states in the constant-field-filled vacuum. It is critical to note that had we separated the sectors along the sign of $Q$, the contents of the theory would be different for $|g|>2$ and unitarity would be violated since the `wrong' levels would be included in the physical half-space.

\noindent{\bf 4. Effective action for $\mathbf{|g|\le 2}$:}\\
\noindent We briefly summarize results for ${|g|\le 2}$~\cite{Labun:2012jf}, as these are needed to understand the novel case of ${|g|>2}$. For constant fields the effective action is manifestly covariant and can be written as a function of the Lorentz-invariant field-like quantities $a,b$
\begin{align}
\!\!
b^2-a^2= \vec B^2-\vec E^2 = \frac{1}{2}F_{\alpha\beta}F^{\alpha\beta}\equiv 2{\cal S}
\;, \quad
(ab)^2 = (\vec E\cdot\vec B)^2 =\left(\frac{1}{8}F^{\alpha\beta}\varepsilon_{\alpha\beta\kappa\lambda}F^{\kappa\lambda}\right)^2 \equiv {\cal P}^2 \label{Pdef},
\end{align}
where $\pm a$ are electric-field-like and $\pm ib$ are the magnetic-field-like eigenvalues of $F^{\alpha\beta}$. $a$ is considered electric-like because $a\to|\vec E|$ on taking the limit $b\to 0$, and similarly $b\to |\vec B|$ in the limit $a\to 0$.

The Schwinger-Fock proper time method~\cite{Schwinger:1951nm} to evaluate the effective action exploits properties of the `squared' 
Dirac equation and thus it can be used to study arbitrary value of $g$. The effective action can be written in the form 
\begin{equation} 
V_{\mathrm{eff}} = \frac{1}{8\pi^2}\int_{0}^{\infty}\frac{du}{u^{3}}\,e^{-i(m^2-i\epsilon)u} F(eau,ebu,\frac{g}{2}).
\label{Veffab}
\end{equation}
For $g=0,2$, the proper time integrand $F (eau,ebu,g)$ was reviewed in Ref.~\cite{Dunne:2004nc}. The generalization throughout the interval $|g|\leq 2$ is accomplished by inserting into Schwinger's Eq.\,(2.33) in the last term a co-factor $g/2$ leading to~\cite{Labun:2012jf}.
\begin{equation} 
\label{VeffabF} 
F(x,y,\frac{g}{2})=\frac{x\cosh(\frac{g}{2} x)}{\sinh x}
\frac{y\cos(\frac{g}{2} y)}{\sin y}-1
,\ \ \left|\frac{g}{2}\right|\le 1 .
\end{equation}
The subtraction $-1$ in~\req{VeffabF} removes the field-independent constant. The logarithmically divergent charge renormalization term is isolated and discussed below. Note that~\req{Veffab} would be divergent for $ |g|> 2$ if~\req{VeffabF} were to be used in this domain.

\noindent{\bf 5. Effective action for $\mathbf{|g|>2}$:}\\
\noindent To extend~\req{VeffabF} to $|g|>2$, we consider in more detail the eigenvalue summation method we introduced above, following the work of Heisenberg and Euler~\cite{Heisenberg:1935qt} and Weisskopf~\cite{Weisskopf}. The mathematical tool used was the L.\,Euler summation formula, leading to the Bernoulli functions $B_{2k}(x)$ and Bernoulli numbers ${\cal B}_{2k}\equiv B_{2k}(0)$. The sum of the Landau energies~\req{spectrum} involves the form $\sum_n f(x+n)$. L.\,Euler developed the technique for such sums, which manifest an integer shift symmetry in the variable $x\to x+n'$~\cite{Euler,Apostol}. Due to this shift symmetry, the Bernoulli functions $B_{2k}(x)$ that arise in the context of L.\,Euler summation of Landau energies $E_n$,~\req{spectrum} are {\it periodic}, given by the Fourier series~\cite{Luo} 
\begin{equation}\label{Bernoullifns}
\tilde B_{2k}(t)=(-1)^{k-1}\frac{(2k)!}{2^{2k-1}}\sum_{n=1}^{\infty}\frac{\cos(2\pi nt)}{(n\pi)^{2k}},
\end{equation}
(here only needed for an even value of index, $2k$). In the unit interval, $0\leq t\leq 1$, the periodic Bernoulli functions are equal to the Bernoulli polynomials, e.g. $\tilde B_{2}(t)=B_2(t)=t^2-t+1/6,\ 0\leq t\leq 1$. Outside the unit interval, the periodic Bernoulli functions~\req{Bernoullifns} $\tilde B_{2k}(t)$ repeat the polynomials' behavior on $0\leq t\leq 1$ in each subsequent period.

Dividing the Landau energies by $2|e\vec B|$ to make the coefficient of $n$ unity, we see that $t\to g/4+1/2$ and hence we recognize that the periodic Bernoulli functions with argument $t=g/4+1/2$ appears in the effective action, arising from the summation of eigenvalues. The explicit representation of the argument of~\req{Veffab} in terms of Bernoulli functions is arrived at employing the analytic transformation of the integrand of~\req{Veffab}~\cite{Muller:1977mm,Cho:2000ei}.
\begin{align}\label{meroexpB} 
&\;F(x,y,\frac{g}{2})
=(x^2\!-\!y^2)\,2\!\sum_{n=1}^{\infty}\frac{\cos n\pi(\frac{g}{2}+1)}{(n\pi)^2} 
\\ \nonumber
&\;
+2\sum_{n=1}^\infty\frac{(-1)^n\cos(\frac{g}{2} n\pi)}{n^2\pi^2}
\Big(\frac{y^4}{y^2-n^2\pi^2} - \frac{x^4}{x^2+n^2\pi^2}\Big)
+ 4x^2y^2\!\!\sum_{n,\ell =1}^\infty
\frac{(-1)^{n+\ell }\cos(\frac{g}{2} n\pi)\cos(\frac{g}{2} \ell \pi)}{(y^2-n^2\pi^2)(x^2+\ell ^2\pi^2)}
\;.
\end{align}
 
In~\req{meroexpB} to assure the necessary periodicity we have introduced, in accordance with~\req{Bernoullifns}, a series of Bernoulli functions with $t=g/4+1/2$. Equation\,(\ref{meroexpB}) agrees exactly with the known expansion~\cite{Cho:2000ei} in the domain of~\req{VeffabF} $|g|\leq 2$ and provides an analytical continuation into the domain $|g|>2$ having the periodicity property of the effective action identified in study of the full set of eigenvalues. Even after the removal of the charge renormalization subtraction term (first term on RHS of~\req{meroexpB}) the finite remainder of the effective action is manifestly periodic in $g$.

Upon performing the proper time integral~\req{Veffab}, each term in~\req{meroexpB} produces a well-defined result for all $g$. The form~\req{meroexpB} is thus an analytic and convergent extension to $|g|>2$ developed using the Euler summation of the eigenvalues~\req{spectrum}. We note that~\req{meroexpB} extends the pure magnetic action based on Weisskopf summation case to arbitrary EM fields via analytical continuation. An independent verification of this continuation was obtained for pure electric fields~\cite{Evans:2022fsu}, and is addressed for field configurations with nonvanishing pseudoscalar in Ref.\,\cite{Evans:2022EB}. This then completes the proof that the here proposed analytical continuation is unique.

\noindent{\bf 6. Nonperturbative in $\mathbf{g}$ renormalization group $\mathbf{\beta}$ function:}\\ 
\noindent The first non constant term on the right hand side of~\req{meroexpB} proportional to $a^2-b^2$ isolates the logarithmically divergent one-loop $\mathcal{O}(\alpha)$ $V_{\mathrm{eff}}$ subtraction required for charge renormalization. The coefficient of this term is related to the $\beta$-function coefficient $b_0$ as is discussed e.g. in section 5.1 in Ref.\,\cite{Dunne:2004nc}. 

We now evaluate the running of the coupling constant $\alpha$ within the $g$-QED loop expansion of the $\mathbf{\beta}$-renormalization function 
\begin{equation}\label{betaf}
\beta \equiv \mu\frac{\partial\alpha}{\partial\mu}, \quad 
\beta(\alpha)=-\frac{b_0}{2\pi}\alpha^2 +\frac{b_1}{8\pi^2}\alpha^3+\ldots \,.
\end{equation}
The first sum in~\req{meroexpB}, for $g=2$, $\sum_{n=1}^\infty 1/(\pi n)^2=1/6$ and implies the value of $b_0=-4/3$, where factor 4 indicates the 4 components of spin-1/2 particle. For arbitrary $g$, $b_0(g)$ is obtained using~\req{Bernoullifns} to identify this sum as $\tilde B_2(g/4+1/2)$. The character of this function is manifest by reconnecting periodic domains of the familiar Bernoulli polynomial $B_2(t)=t^2-t+1/6$ and the resulting $b_0(g)$ coefficient is given in each domain $g\in [g_{k-1},g_k]$ 
\begin{equation}
 b_0=\:-\frac{4}{3}f_{1,0}(g)
=-\frac{4}{3}\left(\frac{3}{8}(g-4k)^2-\frac{1}{2}\right), \label{betafunc}
\end{equation} 
where $f_{1,0}(g)$ is shown in bottom frame of Fig.\,\ref{fig:Veffexpg}. The subscripts of $f_{i,j}$ indicate the powers of the Lorentz invariants in polynomial expansion $f_{i,j}{\cal S}^i {\cal P}^j$ in~\req{meroexpB}. We see in Fig.\,\ref{fig:Veffexpg} that as a function of $g$, the Dirac value $g_{\rm D}=\pm 2$ is an upper cusp point with $f_{1,0}(g)\leq f_{1,0}(2)=1$. For clarity, two periods are shown in Fig.\,\ref{fig:Veffexpg}. 

Note that our result arises in $g$-QED, applying a nonperturbative method in $g$ to one loop expansion. This approach is necessary in order to obtain the behavior of the $\beta$-function for $|g|>2$. At $g=\pm 2$ we find the unexpected cusp. This feature is missing in perturbative consideration of $\beta(g)$ at one loop level which produces the same functional dependence on $g$ as seen in~\req{betafunc} setting $k=0$. As our study shows, a perturbative expansion around $g=0$ in $g$-QED has a finite convergence interval $|g|\le 2$. This was also seen in the Schwinger proper time integral of the effective action.

The following implications for $g$-QED of the properties of the renormalization group coefficient $b_0(g)$ shown in Fig.\,\ref{fig:Veffexpg} are noteworthy:\\
1.) The D-QED expands around $g=\pm 2$, which points are identified as being non-analytic at $g_\mathrm{D }=2$ in the $g$-QED framework.\\
2.) For any value of $g$ not at the cusp $g_\mathrm{D }=2$, the magnitude $|b_0|$ decreases (and thus the speed of `running' decreases) compared to its value at $g=2$. Considering that the coefficient of the magnetic spin term in~\req{EoMg} is dimensionless, no new scale appears in association with $g$. \\
3.) The presence of the cusp in $b_0$ implies that the running coupling of $g$-QED, comprises the cusp as well. \\
4.) A cross check and confirmation of our result for $b_0(g)$ is obtained in perturbative domain considering the limit $g\to 0$ where $b_0(g\to 0)$ differs as expected in sign and the number of degrees of freedom from the known behavior of scalar particle `QED'. 
\\
5.) In the principal domain $|g|\le 2$, the functional dependence on $g$ we find agrees with the result Eqs.\,(53--57) seen in Ref.\,\cite{AngelesMartinez:2011nt}. Specifically, the leading term for large $q^2$ of the vacuum polarization function, evaluated within the framework of $g$-QED is $-\alpha b_0(g)/(2\pi) \ln (-q^2/m^2)$, seen explicitly in Eq.\,(55) of Ref.\,\cite{AngelesMartinez:2011nt}.\\ 
6.) As the above limit shows, for a range of appropriate gyromagnetic moment values $g$ (including $g=0$) $b_0(g)>0$ is {\em possible}. This produces asymptotic freedom behavior for Abelian fermions. The switch between the infrared stable and the asymptotically free behavior occurs in the principal $g$-domain twice, at $g =\pm 2/\sqrt{3}=\pm 1.155$ and continues periodically e.g. for $g=4-2/\sqrt{3}=2.845$. This mechanism of asymptotic freedom generation by $g$-driven sign reversal is implicit in Eq.\,(56) of Ref.\,\cite{AngelesMartinez:2011nt} (valid in principal domain $|g|\le 2$), but the new mechanism allowing Abelian confinement has not been recognized there. The values of $g$ where the sign of the functions $f_{i,j}$ changes is indicated in Fig.\,\ref{fig:Veffexpg}, up to periodic recurrence.

\noindent{\bf 7. Light-light scattering as function of $\mathbf{g}$:}\\ 
\noindent We find that the cusp at $|g|=2$ reappears in the Heisenberg-Euler action, in the light by light scattering. For the general case of both electric and magnetic fields present, using~\req{meroexpB} we find up to fourth order in the fields
\begin{align}\label{Veffexp} 
V_{\mathrm{eff}} \simeq &
\frac{\alpha}{2\pi}\frac{e^2}{45m^4}
\!\left(4f_{2,0}\:{\cal S}^2+7f_{0,2}\:{\cal P}^2\right) 
\;,
\\ \nonumber
f_{2,0}(g)
&=-30\tilde B_4(g/4+1/2)
=-\frac{15(g-4k)^4}{128}
+\frac{15(g-4k)^2}{16}
-\frac{7}{8} 
\;,
\\ \nonumber
f_{0,2}(g)
&=-\frac{60}{7}\left[\tilde B_4\left(\frac g 4 +\frac 1 2\right)
 -3\tilde B_2^2\left(\frac g 4 + \frac 1 2 \right)\right]
 =\frac{15(g-4k)^4}{224}-\frac{1}{14} 
\;,
\end{align}
where both $f_{2,0}$ and $f_{0,2}$ are normalized to $g=2$ values and presented in Fig.\,\ref{fig:Veffexpg}. $f_{0,2}$ includes a product of two Bernoulli functions with cusp and so has a steeper cusp. In general, our finding is that all $f_{i,j}(g)$ for $j>0$ have cusps at $g=2$ whereas all $f_{i,0}(g), i>1$ are continuous and differentiable at $g=2$, being proportional to higher order $>2$ Bernoulli functions that have vanishing derivatives at $g=2$. Thus only coefficients of terms involving powers of the pseudo scalar field invariant ${\cal P}^2=(\vec E\cdot\vec B)^2$ display cusps at $g=2$.

\noindent{\bf 8. Discussion, Conclusions and Outlook:}\\ 
\noindent Difficulties of D-QED as a stand-alone theory have been known for some time, beginning with the work of G. K\"all\'en~\cite{Kallen:1957ib, Kallen:1972pu}, and perturbative-D-QED is believed by many to be semi-convergent only. Exploration of $g\ne 2$ in a renormalizable theory requires the dimension-4 $g$-QED based on the KGP equation. However, $g$-QED has to begin with 8 degrees of freedom and appropriate division into two half-Hilbert spaces is required. Restriction to the usual Dirac-like 4 degrees of freedom is difficult, as a theory with $g\ne 2$ is in general not unitary~\cite{Veltman:1997am}. 

We resolved this problem by proposing a new eigenstate sorting based on the sign of $K$, see~\req{spectrum1}, leading to a self-adjoint theory that retains Poincar\'e symmetry and contains a complete set of particle-antiparticle states, and thus preserves probability in time evolution and analyticity as function of $g$, up to a countable set of singular points. A consequence of this solution is that the Dirac value $g=g_\mathrm{ D}=2$ is a cusp point of the effective action $V_{\mathrm{eff}}$,~\req{Veffab} evaluated in renormalizable $g$-QED approach. 
 
While~\req{VeffabF} is an analytic function of $g$, the integral of~\req{VeffabF} with the proper time weight~\req{Veffab} does not exist for $|g|>2$. Thus a naive extension of HES effective action to $|g|>2$ is not possible. This parallels the observation that the Klein-Gordon-Pauli operator~\req{EoMg} is not self-adjoint for $|g|>2$. We have shown how the eigenstate level crossing can be recognized and states assigned to half-spaces of the full Hilbert space, leading to a natural self-adjoint extension and a valid theoretical $g$-QED framework for $|g|>2$. The cusp and related nonperturbative in $g$ effects arise considering the self-adjoint extension described. The origin of the cusp singularity is in the periodic crossing of eigenenergies in the spectrum of Landau eigenstates seen in upper section of Fig.\,\ref{fig:Veffexpg} showing the quantity $K$, see~\req{spectrum1}. 
 
We have shown cusps at $g=g_{\rm D}$ for two physical quantities computed for arbitrary $g$:\\
$\bullet$ The renormalization group coefficient $b_0$ proportional to function $f_{1,0}$, see Fig.\,\ref{fig:Veffexpg};\\
$\bullet$ The light-by-light scattering in the long-wavelength limit comprising a smooth function $f_{2,0}$, and for the term $(\vec E\cdot\vec B)^2$ the cusp function $f_{0,2}$, see Fig.\,\ref{fig:Veffexpg}. \\
We have checked that these results can be arrived at directly by the method of $\zeta$-function regularization following Weisskopf~\cite{Weisskopf}. Our results agree with earlier perturbative work in the fundamental domain $-2\le g\le 2$: the functional dependence on $g$ is explicit and the same for the vacuum polarization as had been obtained in Ref.\,\cite{AngelesMartinez:2011nt} in Eq.\,(56). We have shown by explicit computation that an expansion around $g=0$ is valid for $|g|\le 2$ only. 

We believe that our results imply that the D-QED expansion around $g=g_\mathrm{D}$ is incomplete at sufficiently high order: Imagine that we partially resume $g-2$ diagrams with Dyson-Schwinger method, finding an effective electron with $g>2$. In the next step we want to compute the vacuum polarization inserts in other $g-2$ diagrams. Attempts in D-QED framework will encounter new divergences as the $g-2$ correction is dimension-5 operator. On the other hand, we can accomplish this task in $g$-QED: we use the non-perturbative in $g$ renormalization group coefficient $b_0$ to characterize the vacuum polarization loop insert and there are no new divergences. However, the result contains the cusp, and thus is different from the finite order perturbative expansion of D-QED.

Our analysis shows how a complete theory of a point-like fermion with $|g|>2$ can be constructed within $g$-QED in order to allow dynamical description of real world spin-1/2 particles. We have obtained the HES effective potential for an elementary particle with gyromagnetic ratio $g\ne 2$ nonperturbatively in $g$, see~\req{Veffab} and~\req{meroexpB}. We demonstrated a cusp as a function of $g$ at the Dirac value $g=g_\mathrm{ D}=2$. We have shown how this cusp enters the $\beta$-function and $(\vec E\cdot \vec B)^{2n}$ terms of light-light scattering. An interesting theoretical consequence is the possibility of asymptotic freedom in an Abelian theory with anomalous magnetic moment originating in the reversal in sign of the renormalization group coefficient $b_0$ for $g$ in specific domains much different from $g=2$.

\section*{Acknowledgment}
We thank S. Evans for review of the manuscript and fruitful discussions.

\let\doi\relax

\end{document}